# Promoting arm movement practice with a novel wheelchair armrest early after stroke: A randomized controlled trial

Sangjoon J. Kim, PhD [1,2], Vicky Chan, PT, DPT[2], Niko Fullmer, BS[3], Emily R. Rosario, PhD[3], Christine Kim, OT[4], Charles Y. Liu, PhD, MD[3,4,5], Marti Comellas, PhD[6], Daniel K. Zondervan, PhD [7], David J. Reinkensmeyer, PhD[2] and An H. Do, MD[8]

[1] Bionics Research Center, Korea Institute of Science and Technology (KIST), Seoul, Republic of Korea.
[2] Department of Mechanical and Aerospace Engineering, University of California-Irvine, Irvine, CA, USA.
[3] Casa Colina Research Institute, Casa Colina Hospital and Centers for Healthcare, Pomona, CA, USA.
[4] Rancho Research Institute, Rancho Los Amigos National Rehabilitation Center, Downey, USA.
[5] USC Neurorestoration Center and Department of Neurosurgery, Los Angeles, CA, USA.
[6] University of Lleida - Polytechnic School, Lleida, Spain.
[7] Flint Rehabilitation Devices, LCC, Irvine, CA, USA.
[8] Department of Neurology, UC Irvine School of Medicine, Irvine, CA, USA.

**Corresponding author(s):**
An H. Do, Associate Professor of the Department of Neurology, UC Irvine School of Medicine, 1001 Health Sciences Rd, Irvine, CA 92617, United States, Phone: +1-(714) 456-7707, Email: and@hs.uci.edu

Word count: 4,029
Number of Figures and tables: 6

**Abstract**

**Background:** Chronic upper extremity (UE) impairment is common after stroke. This study evaluated Boost, a novel wheelchair-mounted rehabilitation device designed to assist individuals in UE motor recovery during inpatient rehabilitation.

**Method:** Thirty-five stroke inpatients were randomized to perform additional UE exercises alongside standard therapy, using either Boost or a therapist-customized booklet for self-practice. Outcomes included the UE Fugl-Meyer (UEFM) Exam, Box and Block Test, Motor Activity Log, Modified Ashworth Scale, shoulder subluxation, and shoulder pain.

**Results:** At baseline, mean days post-stroke were 11.9±4.6 and 13.1±5.9, and UEFM scores were 20.5±10.1 and 21.0±13.5. Intervention durations averaged 11.9±4.0 and 17.2±8.8 days, respectively. Participants in the Boost group completed 3,359±3,137 additional arm movements. No significant between-group differences were found at the three-month follow-up. However, the Boost group showed a trend toward greater UEFM improvement immediately post-intervention (11.8 vs. 6.9 points, $p=0.06$). Importantly, UEFM gains were predicted by the number of Boost exercises performed ($p=0.02$, $R^2=0.34$). Subgroup analysis revealed that patients with less severe impairment (baseline UEFM >21) achieved significantly greater UEFM improvements at discharge with Boost compared to controls (15.8 vs. 7.8 points, $p=0.01$).

**Conclusions:** These findings demonstrate the feasibility of achieving thousands of additional UE practice movements while seated in a wheelchair without direct supervision during subacute rehabilitation. The added movement practice was well tolerated and may offer short-term impairment-reduction benefits, particularly in those with less severe impairment. Larger trials are needed to confirm efficacy, establish optimal dosage, and determine long-term clinical and functional benefits of Boost-assisted therapy.

This study was registered at ClinicalTrials.gov with the title “Comparing Different Rehabilitation Exercise Strategies for Improving Arm Recovery After Stroke (Boost)” in May 2023; registration number: NCT05880940.

**Keywords:** Stroke, rehabilitation, randomized clinical trial, wheelchair mounted device, unsupervised training

## Introduction

It is likely that individuals with upper extremity (UE) impairment after stroke do not engage in sufficient movement practice[1,2] during the early post-stroke period when neuroplasticity is heightened[3–6]. Repetitive UE movement early after stroke is critical for promoting motor recovery, but patients often perform too few voluntary movements during inpatient rehabilitation [7,8]. This limited practice is partly due to muscle weakness and the effort required to move the arm against gravity. To address this gap, we developed a novel, wheelchair-mounted device called Boost (Fig. 1), designed to facilitate a greater amount of early UE motor practice. Boost is a dynamic wheelchair arm support that enables users with UE weakness to repetitively activate their shoulder flexor and elbow extensor muscles—as well as associated proprioceptors—by reaching forward with gravitational support. This design was inspired by two key observations.

First, Feys et al.[9,10] demonstrated the benefits of repeated UE movement in patients with subacute stroke (N=100). Participants who rocked themselves in rocking chairs by repetitively reaching forward to press against a rail (performing 500-1000 UE movements per day) exhibited a significantly greater increase in UE Fugl-Meyer (UEFM) score (+17 points at five-years) compared to a control group that was passively rocked[10]. This finding supports the idea that simple, repeated, active UE movement can yield clinically meaningful benefits.

Second, approximately 70% of stroke inpatients (and nearly 100% of those with severe impairments) spend several hours each day sitting passively in manual wheelchairs, often with their paretic arm resting motionless on an arm rest or trough[11–15]. We sought to leverage this “wheelchair time” for UE movement practice without requiring a transfer to a rocking chair or continuous therapist supervision.

Our previous research established the feasibility and preliminary efficacy of a wheelchair-based UE exercise approach modeled after the Feys et al. rocking chair approach. Initially, we developed a lever attachment for wheelchair wheels, enabling arm movement practice through pushing the lever[16,17]. This intervention reduced arm impairment in a small pilot study with chronic stroke participants[18]. However, users found the device cumbersome to attach and remove. To address this, we designed a lever-drive wheelchair intended as the user’s primary wheelchair, which significantly reduced UE impairment in a pilot randomized controlled trial[19,20]. However, therapists were hesitant to replace standard wheelchairs, highlighting the need for a more compact solution that could be quickly attached to a standard wheelchair.

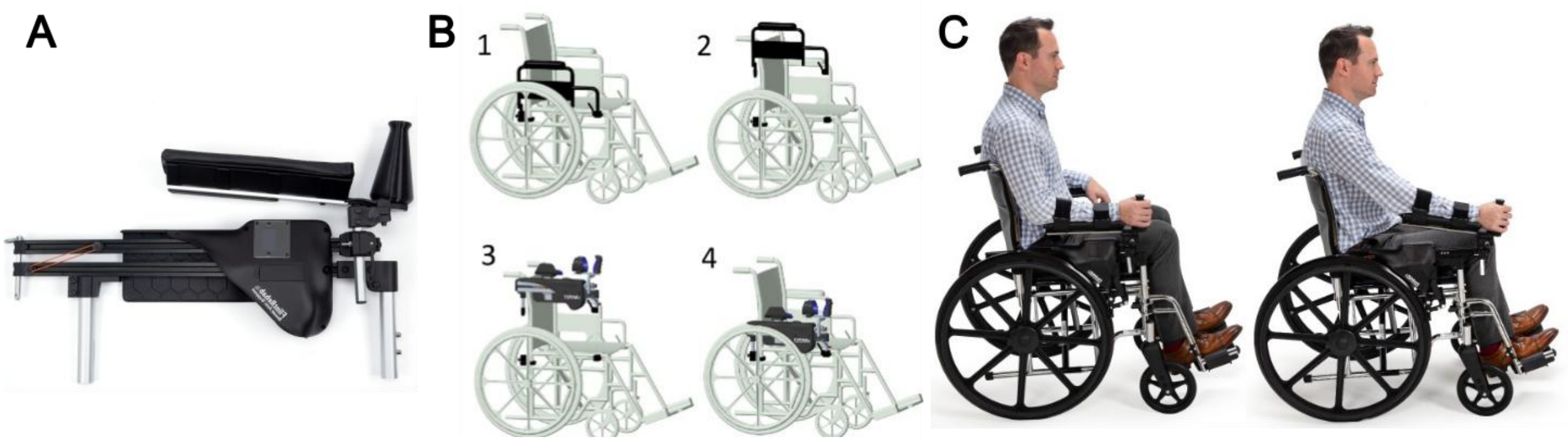

**Figure 1** Description of Boost. (A) Boost is a novel dynamic wheelchair arm support that allows users with weak upper extremities (UE) to repetitively activate their UE by reaching forward in a supported, forward reach pattern while stationary in their wheelchair, (B) A standard wheelchair armrest can be quickly replaced by Boost using the existing armrest slots making it easy to integrate Boost in routine practice (see steps 1-4), and (C) example of performing a stationary forward reaching motion.

Boost was developed in response to this feedback. Boost can substitute for the existing armrest of a conventional wheelchair, securely attaching via the standard armrest slots (Fig. 1B). Once installed, it supports the arm in an ergonomic position, enabling users to perform full-range forward-reaching movements against an adjustable resistance band while the wheelchair remains stationary. Additionally, a built-in digital counter tracks and displays the number of forward arm movements, providing real-time feedback on usage.

This study evaluated the feasibility of using Boost to enhance movement practice during inpatient stroke rehabilitation. Participants who had recently experienced a stroke were randomized to try to practice additional UE exercise with either Boost or a therapist-prescribed exercise booklet. We hypothesized that individuals would be able to use Boost to achieve additional UE movement practice during wheelchair time, and that this practice would reduce UE impairment at discharge and three months post-stroke.

## Method

A multi-site, randomized, assessor-blinded, controlled Phase II trial with parallel design was conducted to assess the potential clinical efficacy of Boost for inpatients with subacute stroke (Fig. 2). The study was registered to ClinicalTrials.gov (NCT05880940). All procedures were approved by the Institutional Review Board (IRB) of the University of California at Irvine, Rancho Los Amigos National Rehabilitation Hospital and Casa Colina Hospital and Center of Healthcare.

*Study design and timeline.* At baseline, eligible participants were stratified into two groups based on upper extremity motor impairment severity (UEFM score 0–21 vs. 22–42) and randomly assigned by the blinded study therapist to either the Boost intervention group or the therapist-guided exercise booklet control group using permuted block randomization. Clinical outcome assessments were conducted at three time points: 1) baseline (within three days of Acute Rehabilitation Unit (ARU) admission), 2) post-therapy (within three days before discharge from the ARU), and 3) three-month follow-up (within three days before or after the date of their stroke plus 90 days) (Fig. 2).

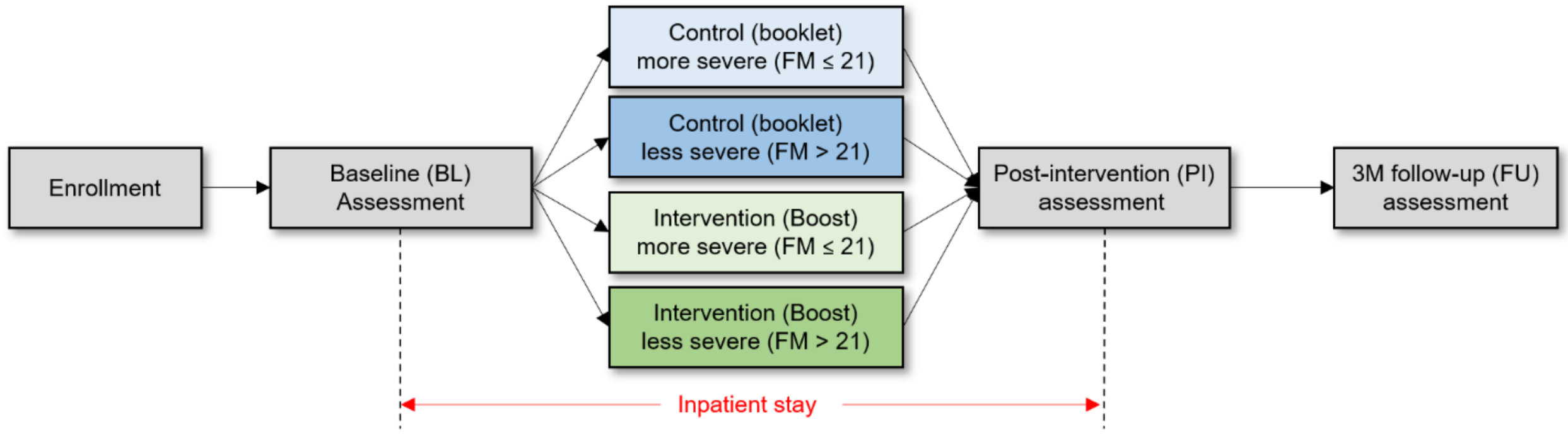

**Figure 2** Overview of the study design.

*Participants.* Recruitment for this study took place between Aug 2023 to November 2024 separately at Rancho Los Amigos National Rehabilitation Center, Casa Colina Hospital and Center for Healthcare and University of California Irvine Health. Inclusion criteria included: (1) age between 18 and 84 years, (2) a history of single or multiple strokes occurring within 3 days to 4 weeks prior to enrollment, (3) admission or acceptance into an ARU, (4) an Upper Extremity Fugl-Meyer (UEFM) Motor Score of less than 42 out of 66, (5) report no moderate to severe shoulder pain during use of the Boost device, defined as a score below 6 on the 10-point Visual Analog Scale, (6) the absence of severe spasticity in the affected upper extremity, indicated by a Modified Ashworth Scale score of less than 4. (7) ARU clinicians were required to confirm that each participant was an appropriate candidate for manual wheelchair use and, finally, (8) participants had to be able to transfer into a wheelchair, with or without assistance, and tolerate sitting in it for a minimum of 30 minutes. Exclusion criteria include individuals with (1) a history of subarachnoid hemorrhage, (2) other neurological or psychological disorders affecting motor function, (3) pregnancy, and (4) significant impairments in vision, language, attention, neglect, or other cognitive functions severe enough to interfere with the safe operation of a wheelchair or the movable wheelchair armrest device were not eligible for the study.

*Intervention (Boost) group.* The intervention group utilized *Boost*, a novel dynamic wheelchair arm support designed to facilitate repetitive upper extremity (UE) activation in individuals with weakness[25–27]. *Boost* enables users to reach forward in an ergonomically friendly, supported, forward reach pattern, either while stationary or while propelling their wheelchair overground. Boost usage is quantified as the number of times the arm passed a magnetic sensor on the track 20 cm from the backstop. This value is automatically logged, recording the number of forward reach exercise repetitions performed. Collected data is securely stored in an encrypted database, with access limited to authorized personnel. Boost is compatible with standard wheelchair armrest slots, allowing for quick interchangeability between conventional commercial armrests and *Boost* (Fig. 1B). *Boost* features two functional modes: “Stationary Mode” and “Overground Mode”. In “Stationary Mode”, users practice a forward-reaching movement with their paretic arm while the wheelchair remains stationary. Resistance of the forward reaching motion can be adjusted by adding or removing resistive bands to customize the level of challenge. In “Overground Mode”, a reel-drive mechanism converts the forward-reaching motion into wheelchair propulsion. *Boost* can also be mechanically locked to function as a conventional armrest. A video demonstrating the use of *Boost* by an individual with severe arm impairment in the subacute phase following a stroke can be found in the supplementary section.

Participants assigned to the intervention group received a wheelchair equipped with *Boost*, with the armrest width (small, medium, large) selected based on arm size, and the height and lateral position adjusted using the quick-adjust mechanism. Therapists provided training to ensure proper use with minimal compensatory movements, such as excessive trunk motion or shoulder hiking. Seat and back cushions were incorporated to promote an upright sitting posture. Once trained, participants were encouraged to use *Boost* independently within the inpatient facility, aiming for at least 30 minutes per day in addition to their standard rehabilitation therapy at the ARU, though usage was not limited to this duration. This daily target was chosen because performing a movement every 3 seconds over 30 minutes would yield approximately 600 practice movements per day—an intensity shown to promote forelimb recovery in rodent models of stroke rehabilitation[28]. Supervising therapists monitored progress and adjusted resistance levels by adding or removing resistive bands to provide an optimal challenge for each participant.

*Control (Therapist-customized exercise booklet) group.* This group followed a therapist-customized exercise program using an exercise booklet tailored to each participant's needs. The booklet was created through a commercially available home exercise program platform (Medbridge), where therapists selected the most appropriate upper extremity (UE) exercises for each individual from the library of exercises. Therapists also provided personal recommendations for exercise repetitions and durations. Participants received training on how to correctly perform their assigned exercises and were encouraged to engage in at least 30 minutes of exercise per day, in addition to their standard rehabilitation therapy at the ARU. However, exercise duration was not strictly limited to this timeframe. Therapists continuously monitored participants' progress and adjusted the difficulty of the exercises as needed to provide an appropriate level of challenge.

*Clinical outcome measures.* The primary outcome measure was the improvement in UE motor function, assessed using the UEFM[21], Secondary outcome measures included the Box and Block Test (BBT)[22], Modified Ashworth Spasticity Scale (MAS)[23], shoulder subluxation distance (SSD), and amount of pain with and without movement. At one study site, each evaluation was conducted with two assessors present, and the same blinded assessor administered all assessments for a given participant across the three time points, ensuring both consistency and blinding. At the other sites, a single blinded assessor conducted all evaluations for both groups at each time point. To ensure standardization, all assessors were certified in the Fugl-Meyer Assessment through BlueCloud prior to the study's start. In addition to the primary and secondary outcome measures, the Motor Activity Log (MAL)[24] was administered at the 3-month follow-up to assess how frequently and effectively participants used their affected arm in daily activities post-discharge. Harms were defined as any adverse events or serious adverse events potentially related to the intervention, including pain or skin irritation. They were assessed through daily therapist monitoring, participant self-report, and review of medical records, and classified by severity and relation to the intervention.

*Data analysis.* We hypothesized that individuals would be able to use Boost to achieve additional upper extremity (UE) movement practice during their inpatient stay. We tested this hypothesis by analyzing the number of movements recorded by the Boost devices. The primary endpoint was the change in UE motor function, assessed by the UEEFM score. We hypothesized that participants who received the Boost intervention would demonstrate significantly greater improvements in UEFM scores compared to the control group ($p < 0.05$, two-sided t-test), without an associated increase in pain or spasticity. Secondary endpoints included changes in the BBT, MAS, SSD, amount of pain with and without movement and MAL score. Additionally, an exploratory endpoint analysis examined outcomes stratified by baseline impairment severity (UEFM≤21 vs. UEFM >21). This subgroup analysis was predefined in the study design to assess whether participants with more severe or less severe impairment derived greater benefits from the intervention.

To assess statistical significance in outcome changes, we first tested for normality using the Shapiro-Wilk test. Depending on the results, we applied either a two-sided Welch's t-test (for normally distributed groups) or a Mann–Whitney U test (for non-normally distributed data). Additionally, we removed participants identified as outliers by examining whether the change in any outcome measure between baseline and post-therapy or baseline and follow-up exceeded 2.5 standard deviations from the mean change across all participants (i.e., both intervention and control groups combined).

## Results

*Recruitment and randomization.* This study initially aimed at recruiting 58 participants based on a previous pilot study[27], however, it was concluded with 35 participants as it reached the end of its planned funding period. Two participants completed the baseline assessment but subsequently discontinued with the study. One participant experienced a second stroke during inpatient rehabilitation and voluntarily withdrew from the study. The other participant had significant cognitive impairments that limited their ability to learn how to use the Boost device. This resulted in 33 participants completing the study. Two participants in the control group were excluded from analysis as outliers. One participant was removed as an outlier because their change in UEFM from baseline to three-month follow-up was 51 points while the mean of the group change was 20 points (SD=12 points). Another participant in the control group was removed as an outlier from analysis because their change in SSD from baseline to three-month follow-up was 4 cm while the mean of the group change was 0.4 cm (SD=1.0 cm). Therefore, data from 31 participants were analyzed (Fig. 3).

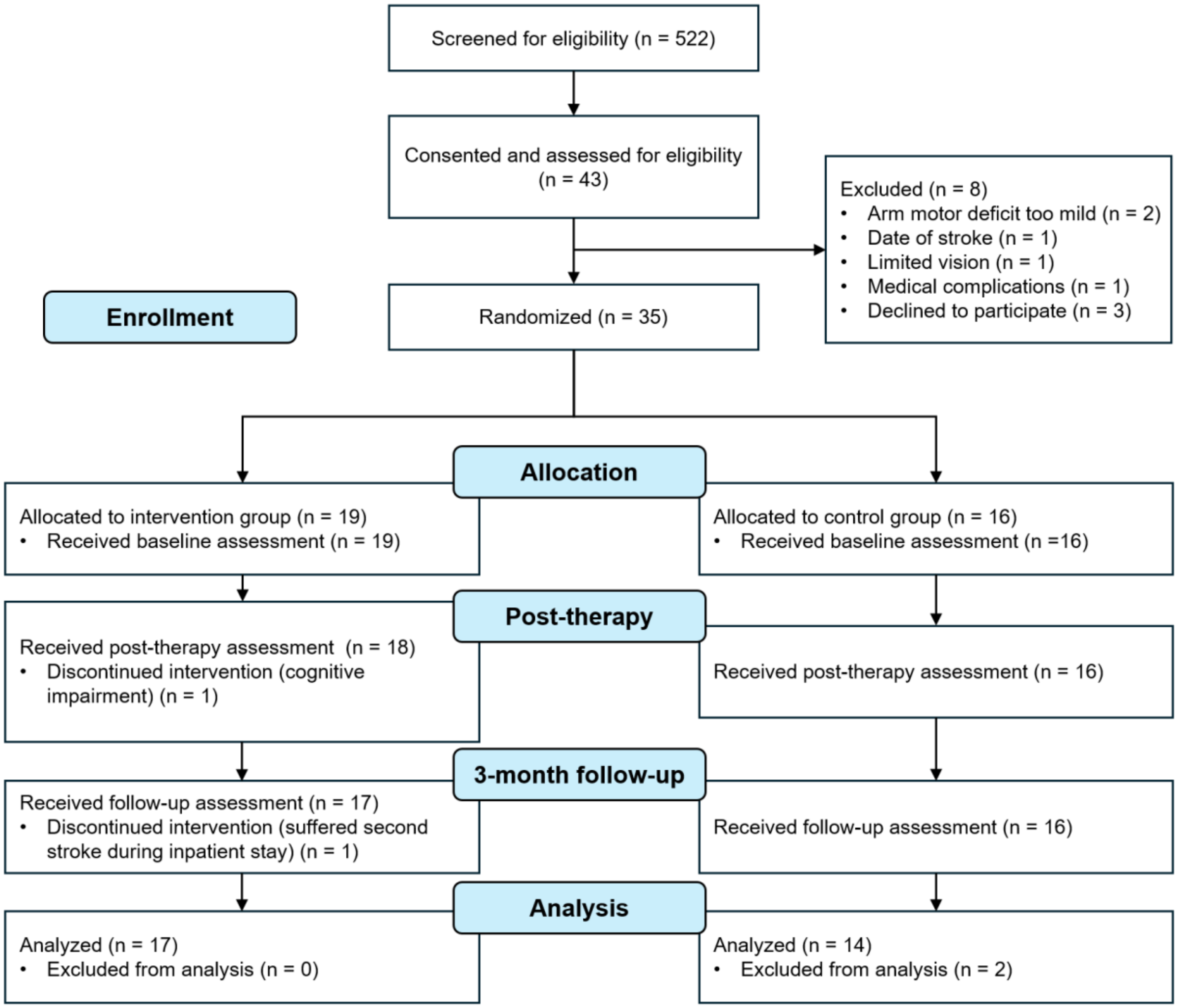

**Figure 3** CONSORT flow diagram. Subjects who did not return for post-therapy assessment were not considered for analysis. Missing data was imputed as described in the statistical methods to maintain group sizes across all analyses.

There were no significant differences in any of the baseline characteristics except the intervention duration (Table 1). The significant difference in the inpatient stay duration was due to a participant in the control group who stayed for 42 days. Removing this individual, the intervention duration was 15.9±6.4 (p = 0.07, two-sided t-test) for the control group.

**Table 1 Baseline characteristics - All participants**

| | Intervention (Boost, n = 17) N (%) | Control (Booklet, n = 14) N (%) | P-Value |
|---|---|---|---|
| **Age (years)** | | | |
| Mean (SD) | 58.7 (15.3) | 59.5 (12.3) | 0.87 [a] |
| Range | 28-79 | 44-85 | |
| **Sex** | | | |
| Male | 7 (41) | 10 (68.8) | 0.25 [b] |
| Female | 10 (59) | 4 (31.2) | |
| **Paretic side** | | | |
| Right | 4 (23.5) | 5 (37.5) | 0.70 [b] |
| Left | 13 (76.5) | 9 (62.5) | |
| **Type of stroke** | | | |
| Ischemic | 14 (82) | 9 (69) | 0.20 [b] |
| Hemorrhagic | 3 (18) | 5 (31) | |
| **Stroke risk factors** | | | |
| High cholesterol | 11 (65) | 8 (50) | 0.67 [b] |
| Hypertension | 14 (82) | 14 (100) | 0.10[b] |
| Diabetes mellitus | 10 (59) | 9 (56) | 0.76 [b] |
| Coronary artery disease | 1 (6) | 1 (12.5) | 0.89 [b] |
| Smoking / Alcoholism | 4 (24) | 2 (18.8) | 0.52 [b] |
| Substance abuse | 2 (12) | 1(6.3) | 0.66 [b] |
| **Time post-stroke (days)** | | | |
| Mean (SD) | 11.9 (4.6) | 13.1 (5.9) | 0.56 [a] |
| Range | 4-19 | 4-27 | |
| **Intervention duration (days)** | | | |
| Mean (SD) | 11.9(4.0) | 17.7 (9.3) | 0.04 [a] |
| Range | 6-20 | 6-42 | |
| **Severity (National Institute of Health Stroke Scale)** | | | |
| Mild (0-7) | 10 (59) | 10 (68.8) | 0.65 [c] |
| Moderate (8-16) | 7 (41) | 4 (31.2) | |
| Severe (>16) | 0 (0) | 0 (0) | |
| **Upper extremity Fugl-Meyer** | | | |
| Mean (SD) | 20.5 (9.8) | 21.0 (12.8) | 0.91 [a] |
| Range | 7-41 | 0-42 | |

SD = standard deviation

[a] Analyzed by Welch’s t-test.

[b] Analyzed by a z-test.

[c] Analyzed by a Chi-Square test.

*Primary Endpoint outcomes.* At the request of therapists at the study sites, participants in this study were restricted to using *Boost* only in “Stationary Mode”. This decision was made out of concern for patient safety, as many participants had moderate to severe impairment and were not appropriate candidates for independent wheelchair propulsion. Individuals in the intervention group completed an average of 3,359 ± 3,137 additional arm movements using Boost in addition to standard therapy. This corresponded to 278 ± 261 additional arm movements per day, calculated by the number of total repetitions divided by the intervention duration.

The intervention group exhibited a trend toward greater improvement in UEFM scores compared to the control group immediately post-intervention (11.8 vs. 6.9 points, $p = 0.06$, Fig. 2). However, this difference diminished at follow-up, with scores of 18.6 vs. 19.2 points ($p = 0.89$). Although not statistically significant, the control group demonstrated a tendency for higher improvement in BBT at the 3-month follow-up (16.0 vs. 22.6 points, $p = 0.29$).

**Table 2 Clinical outcome measures – All participants**

| | Intervention (Boost, n = 17) | Control (Booklet, n = 14) | Difference (95% CI) | P-value |
|---|---|---|---|---|
| | Mean (SD) | Mean (SD) | | |
| **Upper-extremity Fugl-Meyer** | | | | |
| Baseline | 20.5 (9.8) | 21.0 (12.8) | [-9.4, 8.4] | 0.91 [a] |
| Post-intervention | 32.4 (14.6) | 27.9 (14.4) | [-6.6, 15.6] | 0.41 [a] |
| 3-month follow-up | 38.8 (17.0) | 39.8 (15.0) | [-14.5, 12.3] | 0.87 [a] |
| Change in score PI-BL | 11.8 (6.8) | 6.9 (6.9) | [-0.3, 10.2] | 0.06 [a] |
| Change in score FU-BL | 18.6 (11.3) | 19.2 (9.4) | [-9.2, 8.1] | 0.89 [a] |
| **Box and Block test** [b] | | | | |
| Baseline | 3.2 (5.9) | 1.9 (4.7) | | 0.43 [c] |
| Post-intervention | 8.2 (9.4) | 6.7 (9.6) | | 0.62 [c] |
| 3-month follow-up | 16.0 (14.3) | 22.5 (14.9) | | 0.35 [c] |
| **Modified Ashworth Scale Wrist** [c] | | | | |
| Baseline | 0.9 (0.9) | 1.0 (1.1) | | 0.87 [c] |
| Post-intervention | 1.3 (1.1) | 1.0 (0.8) | | 0.44 [c] |
| 3-month follow-up | 1.5 (1.1) | 1.4 (1.0) | [-0.8, 1.0] | 0.76 [a] |
| **Modified Ashworth Scale Elbow** [d] | | | | |
| Baseline | 1.7 (1.1) | 1.7 (1.1) | | 0.98 [c] |
| Post-intervention | 1.7 (1.3) | 1.5 (1.1) | | 0.52 [c] |
| 3-month follow-up | 1.9 (1.5) | 1.9 (1.1) | | 0.74 [c] |
| **Motor Activity Log** [e] | | | | |
| How Well | 1.9 (1.2) | 2.0 (1.4) | [-1.2, 1.0] | 0.87 [a] |
| Amount | 2.0 (1.4) | 2.1 (1.6) | [-1.4, 1.2] | 0.82 [a] |

| | | | |
|---|---|---|---|
| **Shoulder Subluxation Distance Average (cm)** | | | |
| Baseline | 0.6 (0.7) | 0.6 (0.9) | 0.85 [c] |
| Post-intervention | 0.4 (0.8) | 0.2 (0.4) | 0.43 [c] |
| 3-month follow-up | 0.5 (0.8) | 0.3 (0.6) | 0.71 [c] |
| **Amount of pain no movement** | | | |
| Baseline | 0.4 (1.6) | 0.0 (0.0) | 0.40 [c] |
| Post-intervention | 0.4 (1.6) | 0.0 (0.0) | 0.40 [c] |
| 3-month follow-up | 0.8 (1.8) | 0.7 (1.8) | 0.97 [c] |
| **Amount of pain with movement** | | | |
| Baseline | 1.4 (2.0) | 1.5 (2.5) | 0.92 [c] |
| Post-intervention | 1.9 (3.1) | 1.0 (2.0) | 0.36 [c] |
| 3-month follow-up | 3.2 (3.8) | 4.8 (3.2) | 0.38 [c] |

SD = standard deviation; BL = baseline; PI = post-intervention; FU = 3-month follow-up

[a] Analyzed by Welch's t-test.

[b] Reported for paretic side.

[c] Analyzed by Mann-Whitney U test.

[d] Scores were either 1, $1^{+}$, 2, 3, or 4, where $1^{+}$ was assigned the value of 1.5.

[e] Collected at 3-month follow-up.

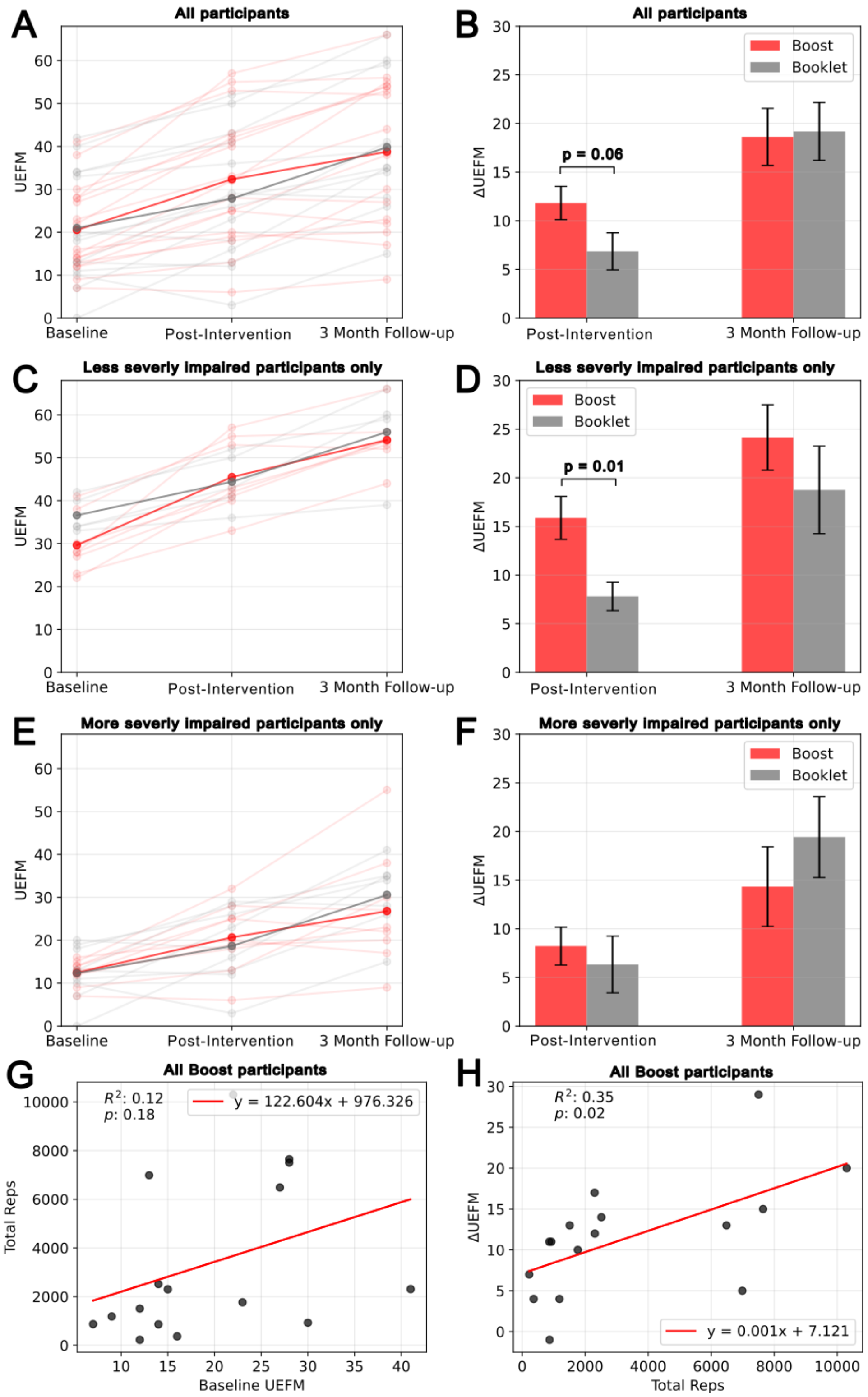

**Figure 4** Comparison of the primary clinical outcome, the UEFM score, for the intervention and control group. (A) UEFM score at each assessment time point

(baseline, post-therapy, and three-month follow-up). (B) the change in UEFM for both groups with respect to baseline UEFM. (C) and (D) show the same thing as (A) and (B) but for the individuals stratified into the less severe arm impairment group at baseline (UEFM >21). (E) and (F) show the same thing as (A) and (B) but for the individuals stratified into the more severely impairment at baseline (UEFM <=21). (G)The relationship between baseline UEFM and the total number of UE exercise repetitions performed using Boost for all participants. (H) The relationship between the total number of UE exercise repetitions performed using Boost and the change in UEFM at the post-therapy evaluation of all participants.

*Secondary endpoint outcomes.* There were no statistical differences in any of the secondary outcome measures.

*Exploratory endpoint outcomes.* At enrollment, we stratified individuals using an UEFM score of 21 into more severe and less severe groups before randomizing in order to balance baseline impairment between groups. Among participants with less severe arm impairment at baseline (UEFM > 21), the Boost group demonstrated significantly greater UEFM improvements at discharge (15.9 vs. 7.8 pts, p = 0.01, Fig. 3). Although not significant, the Boost group exhibited greater improvement in UEFM scores compared to the control group at 3-month follow-up (24.1 vs. 18.8 pts, p = 0.37, Table 3). There was no significant difference between the amount of pain with no movement and with movement for both post-intervention and 3-month follow-up. The primary and secondary outcome measures for the more severe and less severe group are summarized in the supplementary section. During the experiment, the less severe group completed an average of 5,277 ± 3,324 movements using Boost. This corresponded to 440 ± 277 arm movements per day, calculated by the number of total repetitions divided by the intervention duration. Baseline UEFM did predict the total number of Boost exercise repetitions performed (Fig. 3G). However, the total number of Boost exercises performed predicted UEFM score gains post-intervention ($p = 0.02$, $R2 = 0.34$, Fig. 3H).

*Adverse events.* One adverse event occurred during the study. One early participant developed a minor skin abrasion on the forearm caused by the arm trough strap. The abrasion was evaluated and treated by the supervising research physician. In response, padding was added to the arm trough, and the participant wore a cotton forearm sleeve during subsequent use of Boost, which allowed them to continue participation. No abrasions occurred for the rest of the study.

**Discussion**

The core idea of the device tested in this study is to enable individuals with substantial arm weakness early after stroke to stimulate their UE sensory motor system during times that they would otherwise spend sitting passively in their wheelchairs. On average, we found that individuals performed over 3,000 additional arm movements, more than 250 per day (~400 per day for participants with less severe arm impairment), during the average intervention period of 12 days when they had access to Boost on their wheelchair. These additional movements did not increase UE pain, shoulder subluxation, or spasticity, supporting the feasibility and safety of Boost. There was a nearly significantly greater increase in primary outcome, the UEFM score, at the post-therapy evaluation for the Boost group compared to the booklet group, but not in other outcome measures. Examining only the participants who were stratified to the less impaired group (UEFM > 21) at baseline, there was a significantly greater increase in UEFM at the post-therapy evaluation. There were no other significant differences between the other outcome measures between groups. We discuss now these results, including study limitations and directions for future research.

*Evaluating the therapeutic benefit of wheelchair-based UE exercise*

The human motor system retains substantial capacity for plasticity after a stroke, and thus motor impairment of the UE can be reduced with intensive rehabilitation[29–33]. Unfortunately, the intensity of standard UE therapy is likely at least an order of magnitude too low to reach the threshold level necessary to induce substantial plasticity[1,2]. We can infer this from a recent meta-analysis of repetitive reaching exercise in a rodent stroke model[34], which indicated that over 400 reaches per day are necessary to improve reaching function. Standard clinical practice typically achieves about 30 practice movements in a typical therapy session[1,2]. Even two recent, large-scale clinical trials that attempted to implement intense UE training were below this 400-reaches-per-day threshold[7,8], demonstrating the challenge of achieving it in routine practice. Notably, these two trials did not demonstrate a significant effect of the studied intervention, which may be because the putative, required threshold of UE activation was not achieved. In contrast, other recent clinical trials that succeeded in reducing UE impairment by a clinically meaningful amount may have delivered UE activation above this hypothetical threshold[35–37].

The results of the current study showed that it was feasible to achieve ~250 additional UE supported (~400 per day for participants with less severe arm impairment), forward reach practice movements per day during the time that the participants were sitting in their wheelchair. During this arm exercise, participants did not receive continuous supervision from a rehabilitation therapist. Nevertheless, the exercise was safe, as it did not increase UE pain, shoulder subluxation, or spasticity.

While Boost successfully facilitated, semi-autonomous, high-repetition UE practice, the long-term benefits of this practice remain unclear. The Boost group showed a trend toward greater short-term improvement in UEFM scores compared to the control group immediately post-intervention; however, this difference diminished by the 3-month follow-up. One possible explanation is that the additional UE practice accelerated recovery, but continued engagement is necessary to maintain gains. Since both groups received standard rehabilitation therapy, it is also possible that this therapy resulted in comparable benefits by the three month follow-up, reducing the observed differences. When we examined only the participants who were stratified to the less impaired group at baseline, we found a significantly larger increase in UEFM score for the Boost group. Recent studies suggest that residual corticospinal tract (CST) function is critical to promoting UE recovery[38–42]. It may be that these less impaired individuals had the CST resources necessary to better support use-dependent recovery. This observation warrants further investigation in a fully powered Phase III clinical trial.

A higher number of Boost exercise repetitions was significantly associated with greater gains in UEFM score, supporting the role of training intensity in motor recovery. A possible explanation is that individuals with less severe impairment at baseline were able to do more Boost exercises because they had better stamina or could move more quickly. These individuals, in turn, might have been expected to experience larger, spontaneous gains in the UEFM score that were unrelated to their exercise amounts. However, baseline UEFM score did not significantly predict total number of repetitions. This leaves open the possibility that there is a dose-response relationship with Boost, in which case achieving more than the 3,000 movements observed here on average would further improve results.

The marginal reductions in UE impairment we observed, as measured by UEFM, did not translate into corresponding improvements in UE function, as assess by the BBT and MAL. Further, although not significant, the Boost group had lower BBT scores compared to the control group at 3-month follow-up. Some studies suggest that neuroplasticity resources may be competitive, meaning that prioritizing high-repetition arm-reaching exercises could limit progress in fine motor control tasks such as grasping[43]. Future research should investigate whether the gains seen with Boost affect hand function recovery and whether a balanced approach to adding additional arm and hand exercise might improve outcomes across different movement scales.

*Implications, limitations and future directions*

Findings from this study show that wheelchair-based UE exercise devices like Boost can feasibly and safely increase arm movement practice during non-therapy hours, without increasing UE pain, shoulder subluxation, or spasticity. Participants performed thousands of additional arm movements while sitting in their wheelchairs, demonstrating that high-repetition, semi-autonomous practice is possible without constant therapist supervision. However, the overall group outcomes were not significantly different from standard therapy at three-month follow-up, and improvements in impairment did not clearly translate into gains in functional use, as measured by tests like the BBT. This means that any conceptual model for integrating this approach into routine stroke rehabilitation must remain conservative and anchored in the data.

The clearest signal of promise is that participants with less severe arm impairment at baseline—who likely retain greater CST integrity—showed significant short-term gains in UEFM score. This suggests that future research should focus specifically on higher-functioning stroke patients who may achieve greater benefit from additional, semi-autonomous practice, and should deliberately define inclusion criteria—such as minimum UEFM score or measures of residual CST integrity—to target this group more precisely.

A key limitation of this study was the smaller-than-planned sample size (35 instead of the planned 58 participants), which reduced statistical power to detect group differences at follow-up. The longer inpatient stays in the control group and untracked adherence to the exercise booklet may also have affected comparisons. While Boost usage was automatically logged, the lack of standardized protocols for daily dose and resistance progression likely contributed to variability in training intensity. Importantly, the short-term gains in the less impaired subgroup diminished by three months, indicating that added practice may accelerate early recovery but is insufficient to sustain benefits without continued access or reinforcement.

Equally important, the next study should test practical strategies to address why these early effects did not persist. This includes exploring whether a higher total dose of practice, more diverse movement training, or extending Boost use into the home environment can help maintain and build on early gains. Adding structured follow-up sessions, remote monitoring, and motivational support may help ensure that participants continue to engage at an intensity that supports neuroplasticity.

Beyond refining patient selection and dosing, the study should be designed to improve the potential transfer of gains in impairment into functional use of the arm and hand in daily life. This may require developing complementary components that target fine motor skills or grasping tasks, which were not the primary focus here but are critical for activities of daily living. Finally, this study's minor skin irritation incident underscores the importance of incorporating user comfort features, such as adjustable padding, breathable materials, and ergonomic design, especially for individuals with impaired sensation or fragile skin. By aligning the target population, dosing protocols, device design, and follow-up strategies with these lessons learned, future trials will be better positioned to determine whether wheelchair-based practice can deliver sustained, clinically meaningful improvements in upper-extremity recovery.

## Acknowledgements

The authors thank Jeanette Gumarang and Hannah Cone from Casa Colina Hospital and Center of Healthcare and Scott Igtanloc, Lesley Arrendondo from Rancho Los Amigos National Rehabilitation Hospital for helping recruit and manage the clinical trial.

**Declaration of conflicting interests**

The author(s) declare the following potential conflicts of interest with respect to the research, authorship, and/or publication of this article: DKZ and DJR are co-founders of and hold equity in Flint Rehabilitation Devices, a company that is commercializing rehabilitation technology. DKZ is currently employed at Flint, and DJR holds equity and is a scientific board member of Hocoma, a manufacturer of rehabilitation technology. The terms of DJR’s interests have been reviewed by the UC Irvine Conflict of Interest Committee.

**Funding**

The author(s) disclosed receipt of the following financial support for the research, authorship, and/or publication of this article: This work was supported by NIH grant R44HD106850 from the National Center for Medical Rehabilitation Research.

**Supplementary Material**

**Table A Clinical outcome measures – participants with less severe arm impairment at baseline (UEFM > 21)**

| | Intervention (Boost, n = 8) | Control (Booklet, n = 5) | Difference (95% CI) | P-value [a] |
|---|---|---|---|---|
| | Mean (SD) | Mean (SD) | | |
| **Upper-extremity Fugl-Meyer** | | | | |
| Baseline | 29.6 (6.3) | 36.6 (3.7) | [-13.6, -0.4] | 0.06 [a] |
| Post-intervention | 45.5 (7.9) | 44.4 (5.9) | [-8.2, 10.4] | 0.80 [a] |
| 3-month follow-up | 54.1 (6.0) | 56.0 (10.2) | [-19.4, 15.7] | 0.78 [a] |
| Change in score PI-BL | 15.9 (5.8) | 7.8 (2.9) | [2.2, 13.9] | 0.01 [a] |
| Change in score FU-BL | 24.1 (8.2) | 18.8 (7.8) | [-8.2, 19.0] | 0.37 [a] |
| **Box and Blocks test** [b] | | | | |
| Baseline | 6.2 (7.3) | 5.4 (6.5) | [-8.8, 10.5] | 0.85 [a] |
| Post-intervention | 15.5 (8.9) | 14.2 (10.5) | [-13.4, 16.0] | 0.84 [a] |
| 3-month follow-up | 26.4 (9.0) | 34.8 (12.2) | [-29.1, 12.5] | 0.34 [a] |
| **Modified Ashworth Scale Wrist** [c] | | | | |
| Baseline | 0.9 (1.1) | 1.2 (0.7) | | 0.49 [c] |
| Post-intervention | 0.9 (0.9) | 1.1 (0.7) | | 0.82 [c] |
| 3-month follow-up | 1.0 (0.9) | 1.0 (1.2) | | 0.84 [c] |
| **Modified Ashworth Scale Elbow** [d] | | | | |
| Baseline | 1.7 (1.1) | 1.7 (0.6) | [-1.1, 1.1] | 0.98 [a] |
| Post-intervention | 1.4 (1.2) | 1.6 (1.1) | [-1.8, 1.4] | 0.76 [a] |
| 3-month follow-up | 1.0 (1.6) | 1.1 (1.1) | | 0.91 [c] |
| **Motor Activity Log** [e] | | | | |
| How Well | 2.7 (1.2) | 3.2 (1.3) | [-2.8, 1.7] | 0.57 [a] |
| Amount | 2.9 (1.5) | 3.7 (1.1) | [-2.9, 1.2] | 0.37 [a] |
| **Shoulder Subluxation Distance Average (cm)** | | | | |
| Baseline | 0.6 (0.7) | 0.0 (0.0) | | 0.09 [c] |

| | | | |
|---|---|---|---|
| Post-intervention | 0.0 (0.1) | 0.0 (0.0) | 0.53 [c] |
| 3-month follow-up | 0.3 (0.5) | 0.0 (0.0) | 0.32 [c] |
| **Amount of pain no movement** | | | |
| Baseline | 0.0 (0.0) | 0.0 (0.0) | |
| Post-intervention | 0.0 (0.0) | 0.0 (0.0) | |
| 3-month follow-up | 0.0 (0.0) | 0.0 (0.0) | |
| **Amount of pain with movement** | | | |
| Baseline | 1.2 (1.6) | 0.0 (0.0) | 0.16 [c] |
| Post-intervention | 1.4 (2.4) | 0.0 (0.0) | 0.29 [c] |
| 3-month follow-up | 1.3 (2.4) | 1.5 (2.6) | 1.00 [c] |

SD = standard deviation; BL = baseline; PI = post-intervention; FU = 3 month follow-up

[a] Analyzed by Welch's t-test.

[b] Reported for paretic side.

[c] Analyzed by Mann-Whitney U test.

[d] Scores were either 1, $1^+$, 2, 3, or 4, where $1^+$ was assigned the value of 1.5.

[e] Collected at 3-month follow-up.

**Table B Clinical outcome measures – participants with less severe arm impairment at baseline (UEFM <= 21)**

| | Intervention (Boost, n = 9) | Control (Booklet, n = 9) | Difference (95% CI) | P-value [a] |
|---|---|---|---|---|
| | Mean (SD) | Mean (SD) | | |
| **Upper-extremity Fugl-Meyer** | | | | |
| Baseline | 12.4 (2.7) | 12.3 (6.0) | [-5.0, 5.2] | 0.96 [a] |
| Post-intervention | 20.7 (7.5) | 18.7 (8.1) | [-6.3, 10.3] | 0.62 [a] |
| 3-month follow-up | 26.8 (12.6) | 30.6 (7.8) | [-15.6, 8.0] | 0.50 [a] |
| Change in score PI-BL | 8.2 (5.5) | 6.3 (8.3) | [-5.6, 9.4] | 0.60 [a] |
| Change in score FU-BL | 14.3 (11.6) | 19.4 (10.2) | [-17.6, 7.5] | 0.40 [a] |
| **Box and Blocks test** [b] | | | | |
| Baseline | 0.4 (1.3) | 0.0 (0.0) | | 0.37 [c] |
| Post-intervention | 1.7 (2.5) | 2.6 (5.7) | | 0.78 [c] |
| 3-month follow-up | 7.9 (12.2) | 15.6 (11.4) | | 0.26 [c] |
| **Modified Ashworth Scale Wrist** [c] | | | | |
| Baseline | 0.9 (0.7) | 0.9 (1.3) | | 0.67 [c] |
| Post-intervention | 1.7 (1.2) | 1.0 (0.9) | [-0.4, 1.8] | 0.22 [a] |
| 3-month follow-up | 1.9 (1.1) | 1.6 (0.9) | | 0.36 [c] |
| **Modified Ashworth Scale Elbow** [d] | | | | |
| Baseline | 1.7 (1.1) | 1.7 (1.2) | | 1.00 [c] |
| Post-intervention | 2.1 (1.3) | 1.4 (1.1) | [-0.6, 1.9] | 0.27[a] |
| 3-month follow-up | 2.6 (1.1) | 2.3 (0.8) | | 0.36 [c] |
| **Motor Activity Log** [e] | | | | |
| How Well | 1.3 (0.8) | 1.3 (0.8) | [-1.0, 1.0] | 1.00 [a] |
| Amount | 1.2 (0.8) | 1.2 (1.0) | [-1.0, 1.1] | 0.90 [a] |
| **Shoulder Subluxation Distance Average (cm)** | | | | |
| Baseline | 0.7 (0.6) | 1.0 (0.9) | [-1.2, 0.5] | 0.41 [a] |
| Post-intervention | 0.8 (1.0) | 0.3 (0.4) | | 0.33 [c] |
| 3-month follow-up | 0.7 (0.9) | 0.5 (0.7) | | 1.00 [c] |
| **Amount of pain no movement** | | | | |
| Baseline | 0.8 (2.2) | 0.0 (0.0) | | 0.37 [c] |
| Post-intervention | 0.8 (2.2) | 0.0 (0.0) | | 0.37 [c] |
| 3-month follow-up | 1.4 (2.2) | 1.1 (2.1) | | 0.85 [c] |
| **Amount of pain with movement** | | | | |
| Baseline | 1.6 (2.3) | 2.3 (2.8) | | 0.58 [c] |
| Post-intervention | 2.4 (3.5) | 1.6 (2.3) | | 0.61 [c] |
| 3-month follow-up | 4.8 (4.0) | 6.7 (1.5) | [-5.4, 1.5] | 0.24 [a] |

SD = standard deviation; BL = baseline; PI = post-intervention; FU = 3 month follow-up

[a] Analyzed by Welch's t-test.

[b] Reported for paretic side.

[c] Analyzed by Mann-Whitney U test.

[d] Scores were either 1, $1^+$, 2, 3, or 4, where $1^+$ was assigned the value of 1.5.

[e] Collected at 3-month follow-up.